\documentclass[apl,prl,twocolumn,groupedaddress]{revtex4}
\usepackage{graphicx}
\usepackage{calc}
\usepackage{pifont}
\usepackage{amssymb}
\usepackage{epsfig}
\usepackage{amssymb}
\usepackage{amsmath}
\usepackage{color}
\DeclareMathAlphabet{\mathitb}{OT1}{cmr}{bx}{sl}
\begin{document}

\renewcommand{\thefootnote}{\fnsymbol{footnote}}
\title{Universal Scaling of Nonequilibrium Transport in the Kondo Regime of Single Molecule Devices}
\author{G. D. Scott$^1$}
\email{gavin.scott@rice.edu}
\author{Z. K. Keane$^1$}
\author{J. W. Ciszek$^{2,3}$}
\author{J. M. Tour$^3$}
\author{D. Natelson$^{1,4}$}
\affiliation{$^1$Department of Physics and Astronomy, $^3$Department of Chemistry,
$^4$ Department of Electrical and Computer Engineering, Rice University, 6100 Main St., Houston, TX 77005\\
$^{2}$Department of Chemistry, Loyola University Chicago, 1068 W. Sheridan Rd., Chicago, IL 60626}

\date{\today}

 \begin{abstract}

Scaling laws and universality are often associated with systems exhibiting
emergent phenomena possessing a characteristic energy scale.  We
report nonequilibrium transport measurements on two different types of single-molecule
transistor (SMT) devices in the Kondo regime.  The conductance at low
bias and temperature adheres to a scaling function characterized by
two parameters.  This result, analogous to that reported recently in
semiconductor dots with Kondo temperatures two orders of magnitude
lower, demonstrates the universality of this scaling form.  We compare the
extracted values of the scaling coefficients to previous
experimental and theoretical results.
\end{abstract}

\maketitle

Scaling theories, most notably of the power law form,
characterize a diverse array of naturally occurring phenomena.
Seemingly unrelated systems with widely differing microscopic details
can be described by relations with identical scaling exponents when
the underlying dynamics are similar.  This property, known as
universality, is relevant when considering
systems driven out of equilibrium by some perturbation.
In a system with an emergent characteristic energy scale,
examining the system response as a function of the perturbation
scaled relative to that energy can reveal such universality.

The many-body Kondo state in quantum dots, as observed in a
variety of microscopic
implementations\cite{Goldhaber1998a,Cronenwett1998,Nygard2000,JPark2002,Liang2002,Yu2004b},
is an example of such a system.  In these devices itinerant electrons
in source and drain leads are coupled via tunneling barriers to a
single magnetic impurity.  At zero source-drain bias ($V$), below
an emergent Kondo energy scale, $k_{\mathrm B}T_{\mathrm K}$, the local moment
of the impurity is screened by the conduction electrons.  As a result
of the screening process the (zero-bias) conductance, $G$, is enhanced
at low temperatures ($T < T_{\mathrm K}$).  It has been documented that
$G(T,V \approx 0)$ in the equilibrium Kondo regime is well
described by a universal temperature
dependence.\cite{Goldhaber1998b,Wiel2000,Yu2004c}
A nonzero applied
bias ($|V| > 0$) drives the system into the nonequilibrium Kondo
regime.\cite{Wingreen1994,Costi1994,Konik2002}  $G(T,V)$ exhibits
a resonance peak centered at $V=0$.
Theoretical treatments
of idealized quantum dots have argued that $G(T,V)$ in the
single channel Kondo state may be described by an analogous
function in bias voltage and temperature.  However, there is
discussion regarding the energy scales and order to which this
scaling will hold as well as the number of system specific
coefficients required, their expected values, and the
universality of these
coefficients.\cite{Schiller1995,Majumdar1998,Kaminski2000,Rosch2001,Oguri2005,Doyon2006}
Recent experiments by Grobis {\em et al.}\cite{Grobis2008} have shown
that $G(T,V)$ measured in a single channel GaAs quantum dot
($T_{\mathrm K} \approx 0.3$~K) in the nonequilibrium regime is well described
by a universal scaling function with two scaling parameters.

In this Letter we test the universality of these results, applying the
analogous analytical approach to 29 molecule-based devices with Kondo
temperatures ranging from 35~K to 155~K.  We find that the
conductances of SMTs containing either
C$_{60}$\cite{Yu2004b,Pasupathy2004,Parks2007} or a transition metal
complex\cite{Yu2005} are accurately described by the same scaled parameters
as the GaAs dot\cite{Grobis2008}.  We confirm the
quadratic voltage and temperature dependence of $G$ in the low energy
limit.  The values of the extracted scaling coefficients are quite
consistent throughout the ensemble.  We discuss possible explanations
for the systematic differences between our coefficients and those
inferred from previous experiments\cite{Grobis2008} and theoretical
model predictions.

\begin{figure}[!t]
\begin{center}
\includegraphics [scale=.39]{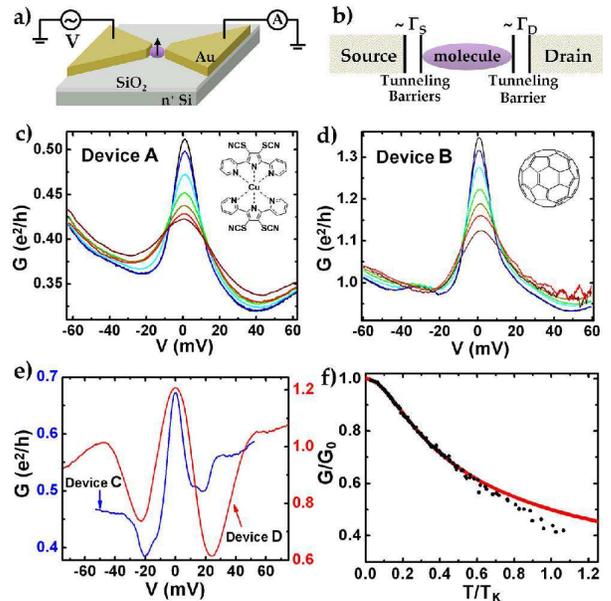}
\end{center}
\vspace{-7mm}
\caption{(color online).  (a) Schematic of SMT measurement setup.  (b) Transport path for conduction electrons in SMT device.  Traces of $dI/dV$ vs. $V$ for several temperatures (c) from 1.8 K(black) to 53 K(maroon) for device \textbf{A}, using complex \textbf{1}, and (d) from 1.8 K(black) to 46 K(maroon) for device \textbf{B}, using C$_{60}$.  Amplitude of Kondo resonance decreases with increasing temperature.  Insets: Schematic of (c) complex \textbf{1} and (d) C$_{60}$.  (e) $dI/dV$ vs $V$ at T = 1.8 K for two samples, demonstrating the disparate Kondo temperatures associated with different devices.  Device \textbf{C} (blue) contains a C$_{60}$ molecule and \textbf{D} (red) contains complex \textbf{1}.  Note that the respective background contributions to the measured conductance have not been subtracted from the plots in (c), (d), and (e).  (f) Fit of equilibrium conductance versus temperature for devices \textbf{A}, \textbf{B}, \textbf{C}, and \textbf{D}, using equation \ref{GEK}, yielding $T_{\mathrm K} \approx$ 52~K, 43~K, 35~K, and 105~K, respectively.}
\label{traces}
\end{figure}

Sample fabrication begins with e-beam lithographic patterning of gold
nanoconstrictions with minimum widths between 90 and 120~nm, in arrays
on an $n^+$ Si substrate with a 200~nm SiO$_2$ insulating layer.
Immediately prior to being placed in a cryostat, the chip is cleaned
and then treated with the molecules that are to be utilized as
the active elements, as described previously\cite{Yu2004b,Yu2005}.
Each device in an array can be contacted individually using an
Attocube Systems cryogenic probe station with a base temperature
$\sim$~1.8~K.  Electromigration in a cryogenic environment is used to
create a small gap in the nanowire, ideally allowing a molecule to
bridge the broken ends of the wire that will subsequently function as
source and drain electrodes (Fig.\ref{traces}a and
\ref{traces}b).\cite{HPark1999} C$_{60}$ and a Cu-containing transition metal
complex (bis(2,5-di-[2]pyridyl-3,4-dithiocyanto-pyrrolate)Cu(II))
(``complex \textbf{1}''), were independently utilized as the active
elements of our SMTs.  Both
C$_{60}$\cite{Yu2004b,Pasupathy2004,Parks2007} and complex
\textbf{1}\cite{Yu2005,Natelson2006} have been successfully used as
active elements in past investigations of breakjunction devices.

Differential conductance, $dI/dV$, was measured as a function of $V$
using quasi-DC lock-in techniques with an rms excitation voltage of
0.75~mV at 19.149~Hz.  Post-electromigration transport properties were
consistent with previous reported frequencies of
occurrence.\cite{Natelson2006} A pronounced zero-bias conductance peak
was observed in approximately 20\% of devices, approximately a third
of which were sufficiently stable at relevant biases and temperatures
to enable a complete analysis.  The
prototypical behavior of the single channel spin-$\frac{1}{2}$ Kondo
effect is exhibited by a decreasing amplitude and
broadening of the Kondo resonance as the temperature is increased.
The temperature dependence of the amplitude
has been well described by an empirical form\cite{Goldhaber1998b} derived
from a fit to the renormalization group analysis.\cite{Costi1994}
\begin{equation}
G_{\mathrm {EK}}(T,0) = G_{\mathrm 0}(1 + (2^{1/s} - 1)(T/T_{\mathrm K})^2)^{-s} + G_{\mathrm b},
\label{GEK}
\end{equation}
where $s = 0.22$ for a spin-$\frac{1}{2}$ impurity.  The total conductance comprises the Kondo resonance and a background contribution, $G_{\mathrm b}$, approximated as linear in applied bias, found via a fit to $dI/dV$ vs $V$ over a bias range that incorporates the conductance minima outside the Kondo resonance.  A constant is subtracted from the fit to ensure that $G(T,V) - G_{\mathrm b} \geq 0$ for all T and V.  From Eq.~(\ref{GEK}), $G_{\mathrm 0}$ is the height of the resonant peak, in the limit that $T\rightarrow 0$~K:  $G_{\mathrm 0} = G(0,0) - G_{\mathrm b}$.  Similarly $T_{\mathrm K}$ is the temperature at which the Kondo resonance equilibrium conductance is equal to half its zero-temperature value:  $G(T_{\mathrm K},0) - G_{\mathrm b} = G_{\mathrm 0}/2$.  From this description we extract values of the parameters $T_{\mathrm K}$ and $G_{\mathrm 0}$ (Fig.~\ref{traces}f).  As $T$ increases, deviations from Eq.~(\ref{GEK}) are expected as higher order transport processes become relevant.  Our data is in good agreement with this form up to $T/T_{\mathrm K} = 0.4~-~0.75$, depending on the device.  Further references to $G(T,V)$ denote conductance with the background component removed.

Of the 29 devices analyzed over the full temperature range, 19
incorporated complex \textbf{1} and 10 utilized C$_{60}$, with
$T_{\mathrm K}$ ranging from 34~K to 155~K.  While the C$_{60}$-based devices
had a lower $T_{\mathrm K}$ on average ($T_{\mathrm {K,ave}} \approx 57$~K for
C$_{60}$-based devices compared to $T_{\mathrm {K,ave}} \approx 70$~K for
devices utilizing complex \textbf{1}), they were typically less stable
with increasing temperature and bias.  The enhanced stability of devices made with complex \textbf{1} is likely due to
their thiocyanate end groups which can covalently bond to the gold
electrodes.  Irreversible changes in device configuration at higher
temperatures limited our measurements to the region of $T \lesssim
T_{\mathrm K}/2$ for the majority of devices.

If the Kondo resonance scales universally with
respect to bias and temperature, then we can assume that the scaling
function may be approximated by a series expansion.
We test this idea by fitting
the nonequilibrium conductance to the equation
$G(T,V) \approx G_0 - \tilde{c}_T(k_{\mathrm B}T)^{P_T} - \tilde{c}_V(eV)^{P_V}$.
$P_V$ and $P_T$ are the scaling exponents for bias and temperature,
while $\tilde{c}_V$ and $\tilde{c}_T$ are expansion coefficients.  We
do not presume the theoretically predicted quadratic form of the
scaling exponents and instead estimate $P_V$
and $P_T$ independently.  To extract $P_V$ we fit $G(T,0) -
G(T,V)$ to a power law form in the bias range $|{eV}| \leq k_{\mathrm B}T_{\mathrm K}/2$
for each trace in which $T \leq T_{\mathrm K}/5$.  To extract $P_T$ this process
is repeated for $T \leq T_{\mathrm K}/5$ at fixed values of $V$, again in the
range $|{eV}| \leq k_{\mathrm B}T_{\mathrm K}/2$.  We limit the $T$ and $V$ ranges because the nonequilibrium conductance is
expected to deviate from universal scaling as $k_{\mathrm B}T$ and $eV$
approach $k_{\mathrm B}T_{\mathrm K}$.  Averaged
over all devices, we find $P_V = 1.97 \pm 0.047$ and $P_T = 2.07 \pm
0.18$.  Theoretical treatments have found the leading order corrections to the
conductance are quadratic in temperature and bias.\cite{Schiller1995,Nozieres1974,Pustilnik2004}
The scaling exponents we find for both T and V are thus very near the predicted value
of 2.  Henceforth we use $P_V$ and $P_T$ equal to 2.  The error bars for
all extracted parameters in this study represent 95\% confidence intervals.

Following the approach of Grobis {\em et al.}\cite{Grobis2008},
we first obtain a scaling function of normalized temperature,
$T/T_{\mathrm K}$, and bias, $eV/k_{\mathrm B}T_{\mathrm K}$ describing the nonequilibrium
conductance at low energies.  In the context of
the Anderson impurity model\cite{Costi1994,Anderson1961}, the Kondo
resonance arises from the local density of states\cite{Nagaoka2002} of a magnetic impurity with charging
energy $E_{\mathrm c}$ which is coupled to the electron bath of the leads; in a dot
in the Kondo regime this is proportional to the conductance and may be
expanded in the low energy limit as\cite{Grobis2008}
\begin{equation}
\footnotesize{G(T,V) = G_{\mathrm {EK}}(T,0)\left(1 - \frac{c_T\alpha}{1 + c_T\left(\frac{\gamma}{\alpha} - 1\right)\left(\frac{T}{T_{\mathrm K}}\right)^2}\left(\frac{eV}{k_{\mathrm B}T_{\mathrm K}}\right)^2\right)},
\label{GTV1}
\end{equation}
where $\alpha$ and $\gamma$ are the scaling coefficients we intend to
extract.  The constant $c_T \approx 4.92$.  This is set by the low
temperature expansion of Eq.~(\ref{GEK}) in which $G_{\mathrm {EK}}(T,0) \approx
G_{\mathrm 0}(1 - c_T(T/T_{\mathrm K})^2)$.  The
constants $\alpha$, $\gamma$, and $c_T$ are arranged in this fashion
so that Eq.~(\ref{GTV1}) can be expanded at low temperature to a
simplified power law form.  Here, $\alpha$ and $\gamma$ are defined such
that they are independent of the definition of $T_{\mathrm K}$.
Eq.~(\ref{GTV1}) in the small $T$ limit is of the form\cite{Schiller1995,Majumdar1998}
\begin{equation}
\footnotesize{\frac{G(T,0) - G(T,V)}{c_TG_{\mathrm 0}} \approx \alpha\left(\frac{eV}{k_{\mathrm B}T_{\mathrm K}}\right)^2 - c_T\gamma\left(\frac{T}{T_{\mathrm K}}\right)^2\left(\frac{eV}{k_{\mathrm B}T_{\mathrm K}}\right)^2}.
\label{GTV2}
\end{equation}
In this expression $\alpha$ represents the $T=0$ curvature of the
Kondo resonance with respect to $V$ while $\gamma$ describes the
rate at which the resonance peak broadens and decreases in amplitude
with increasing temperature.

\begin{figure}[!t]
\begin{center}
\vspace{-9mm}\includegraphics [scale = .8]{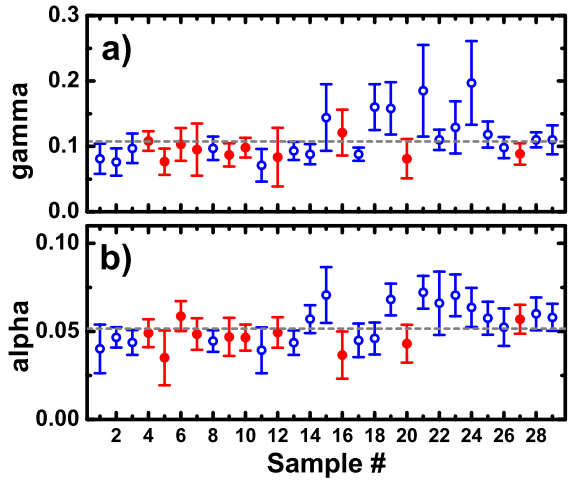}
\end{center}
 \hspace\fill\vspace{-12mm} \caption{(color online).  Extracted values of the scaling coefficient (a) $\alpha$ and (b) $\gamma$ for each device.  Open blue circles correspond to devices with complex \textbf{1}.  Filled red circles correspond to devices with C$_{60}$.  Dashed gray line indicates average value.  Samples are numbered in order of increasing $T_{\mathrm K}$.}
\label{ScalingCoeff}
\end{figure}

Figure \ref{ScalingCoeff} shows the values of $\alpha$ and $\gamma$ extracted
for each sample fitted over the range $T < T_{\mathrm K}/5$ and $eV < k_{\mathrm
  B}T_{\mathrm K}/2$.  We calculate average values of $\alpha = 0.051 \pm
0.01$ and $\gamma = 0.107 \pm 0.027$.  Devices with complex
\textbf{1} had a roughly 10\% larger average $\alpha$ and 18\% larger
average $\gamma$ compared to C$_{60}$-based devices.  This is due
to a few samples containing complex \textbf{1} which possess exceptionally
large values of $\alpha$ and $\gamma$.  These samples may originate
from conduction occurring closer to the mixed valence regime, leading
to non-Kondo processes preventing an accurate description of the
conductance by a universal scaling function.  Beyond this minor
difference, the extracted coefficients are essentially
molecule-independent.

\begin{figure}[!b]
\begin{center}
\includegraphics [width=8cm]{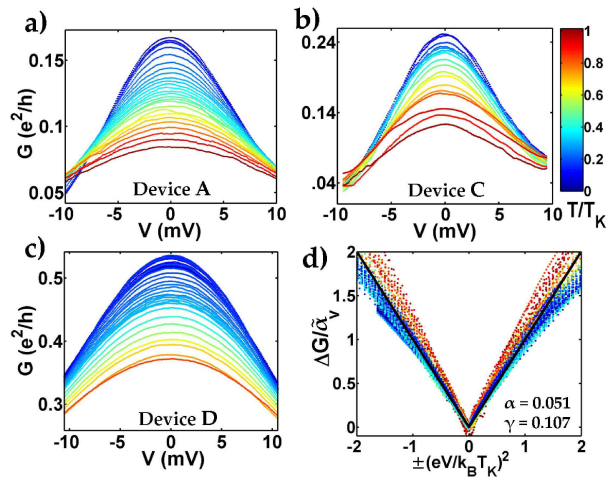}
\end{center}
\hspace\fill \vspace{-9mm} \caption{(color online).  Conductance as a function of $V$ for (a),(b) $T/T_{\mathrm K} \lesssim 1.0$ and (c) $T/T_{\mathrm K} \lesssim .65$, corresponding to devices \textbf{A}, \textbf{C}, and \textbf{D} from Figs. \ref{traces}c and \ref{traces}e, respectively.  $G_{\mathrm b}$ has been subtracted off.  (d) Scaled conductance, $\Delta{G}/{\tilde{\alpha}_V}$, versus $(eV/k_{\mathrm B}T_{\mathrm K})^2$ for devices \textbf{A}, \textbf{B}, \textbf{C}, and \textbf{D}, where $\Delta{G} = (1-G(T,V)/G(T,0))$.  The data from (a), (b), and (c) are plotted here, along with the analogous data set from device \textbf{B}, using the average extracted values of the scaling coefficients.  The solid black line represents the associated universal curve.  The colorbar in (b) pertains to all 4 figures.}
\label{ScaledData}
\end{figure}

Transport measurements ($G(T,V)-G_{\mathrm b}$ vs. $V$) in devices \textbf{A}, \textbf{C}, and \textbf{D} from Fig. \ref{traces} are shown in Figs.~\ref{ScaledData}a \ref{ScaledData}b, and \ref{ScaledData}c, respectively.  In Fig. \ref{ScaledData}d we plot scaled conductance,
$(1-G(T,V)/G(T,0))/{\tilde{\alpha}_V}$, versus $(eV/k_{\mathrm B}T_{\mathrm K})^2$ for the data in these figures, along with the equivalent set of traces from device \textbf{B}, for $(eV/k_{\mathrm B}T_{\mathrm K})^2 \leq 2$ and all available temperatures in the range $T/T_{\mathrm K} \lesssim 1.0$, using the average extracted values of $\alpha$ and $\gamma$.  We define ${\tilde\alpha}_V \equiv c_T\alpha/(1 + c_T\left(\frac{\gamma}{\alpha} - 1\right)({T}/T_{\mathrm K})^2)$. This plot provides a useful qualitative tool indicating the degree to which transport in the Kondo regime conforms to the predicted scaling function.  Data from both devices \textbf{A} and \textbf{B} adhere to the scaling form for $(eV/k_{\mathrm B}T_{\mathrm K})^2 \lesssim 0.6$ and for $T/T_{\mathrm K} \lesssim 0.75$, above which deviations exceed the confidence levels.  For devices \textbf{C/D} the data conform to the scaling function up to $(eV/k_{\mathrm B}T_{\mathrm K})^2 \lesssim 0.5/0.55$ and $T/T_{\mathrm K} \lesssim 0.55/0.45$.  
When data for each device is fit to find that device's particular optimal values of $\alpha$ and $\gamma$, the goodness of fit is essentially the same for every device.
If we assume the premise of universal scaling such that a single (averaged) $\alpha$ and $\gamma$ should describe \textit{all} the devices, data from every device with extracted values of $\alpha$ and $\gamma$ that lie within the average error bars collapse onto the scaling curve (black line in Fig.~\ref{ScaledData}d) up to at least $(eV/k_{\mathrm B}T_{\mathrm K})^2 \lesssim 0.5$ and for $T/T_{\mathrm K} \lesssim 0.4$.

While the conductance in these devices follows the
expected\cite{Schiller1995,Majumdar1998} scaling form, we consider the
purported universality of the coefficients $\alpha$ and
$\gamma$ by examining the level of accord with other experimental data and
theoretical expectations.  The values extracted from selected
experimental measurements of nonequilibrium transport through a GaAs
dot ($T_{\mathrm K} \approx 0.3$~K) with reasonably symmetric coupling (2:1)
indicate $0.1 \lesssim \alpha \lesssim 0.15$ and $\gamma \approx
0.5$.\cite{Grobis2008}, which deviate from our extracted values beyond the statistical uncertainties in the parameters.
Theoretical treatments
based on the Anderson\cite{Costi1994,Konik2002,Oguri2005} and Kondo\cite{Schiller1995,Majumdar1998,Doyon2006,Pustilnik2004} models focus on defining $\alpha$,
which is predicted to take on universal values in the limits of
symmetric or strongly asymmetric device coupling.  Predictions using
various approaches are in the range $0.1 \lesssim \alpha \lesssim 0.3$,
depending on the perturbation method used.  Our smaller extracted values of $\alpha$ imply that the
Kondo resonances that we observe evolve more slowly than expected
with bias voltage given the $T_{\mathrm K}$ values inferred from their
equilibrium temperature dependencies.

There are a number of possible explanations for this systematic difference in
$\alpha$.  The relative asymmetry of the source and drain coupling may be a relevant issue.
The total level broadening $\Gamma = {\Gamma}_{\mathrm S} +
{\Gamma}_{\mathrm D}$, where ${\Gamma}_{\mathrm S}$ and ${\Gamma}_{\mathrm D}$ are contributions of
the respective tunneling barriers as established by the overlap
between the local moment and the conduction electron states of the
source and drain electrodes, respectively.  In a SMT these values are
exponentially sensitive to the precise molecule-electrode
configuration and will generally be different for every device.  We
can infer the asymmetry of the coupling from the magnitude of the
Kondo resonance as $T \rightarrow 0$, predicted to
be $(2e^2/h)(4\Gamma_{\mathrm S}\Gamma_{\mathrm D})/(\Gamma_{\mathrm S}+\Gamma_{\mathrm D})^2$.\cite{Beenakker1991}
For the SMTs studied here we calculate $\Gamma_{\mathrm S}$:$\Gamma_{\mathrm D}$ ratios
between 5.2:1 and 182:1 with a median ratio of approximately 17:1.  No
systematic difference in coefficient values was found for SMTs with
increasing coupling asymmetry ratios.  There were also no clear
correlations found between scaling coefficient values (and thus deviations from the scaling curve) and values of
$T_{\mathrm K}$, $G_{\mathrm 0}$, or $G_{\mathrm b}$.

It is also possible that the molecular character of the devices is
relevant to the systematic difference in $\alpha$.  The essentially
identical observations in C$_{60}$ and
complex {\bf 1} devices suggest that there is likely a single
explanation that does not depend in detail on molecular structure.
This argues against higher-spin Kondo states or orbital degeneracies,
since such physics would differ significantly between molecule types.
Most theoretical treatments assume extremely large (or infinite)
on-site repulsions relative to the single-particle level spacing, a
relationship that is not necessarily valid in SMTs.  Furthermore,
molecular devices have vibrational degrees of
freedom.\cite{Park2000,Yu2004c} These modes have been
suggested\cite{Cornaglia2007,Elste2008} as relevant to the Kondo
regime in SMTs, particularly with respect to the unusually weak gate
dependence of the Kondo resonance.\cite{Yu2005} In addition to
molecule-specific intramolecular vibrational modes, SMTs possess
relatively generic center-of-mass vibrational modes\cite{Park2000} at low energies
($\sim 5-10$~meV) that may be relevant across many different molecule
types.  If the temperature dependence of the Kondo resonance in SMTs
is modified by vibrational effects, our extraction of $T_{\mathrm K}$
values via fitting the temperature dependence may lead to a different
effective $\alpha$ than in the bare Kondo system.

We have measured conductance through SMTs in the Kondo regime and
found that nonequilibrium transport at low energies is well described
by a scaling function of the same functional form as that used for a
GaAs quantum dot in the Kondo regime.  A quadratic power-law in
temperature and bias characterized by two scaling coefficients
accurately reflects the evolution of the Kondo resonance for SMTs made
with either C$_{60}$ or complex \textbf{1}.  The extracted values of
the scaling coefficients deviate systematically from those in the GaAs
case and from theoretical predictions based on idealized models.
These deviations highlight the need for further experimental and
theoretical investigations of molecular devices and for an improved
understanding of the limitations of universality of Kondo physics
in realistic nanostructures.

Thanks to Stefan Kirchner, Qimiao Si, and Mike Grobis for useful discussions.
DN acknowledges support from NSF CAREER award DMR-0347253 and the David
and Lucille Packard Foundation. GDS acknowledges the support of the
W. M. Keck Program in Quantum Materials at Rice University.  JMT acknowledges support from DARPA.

\end{document}